# CONFRONTATION OF THE LAUER AND POSTMAN CLUSTER VELOCITY FIELD WITH MODELS

MICHAEL A. STRAUSS[1], RENYUE CEN[2], AND JEREMIAH P. OSTRIKER[2]
[1] *Institute for Advanced Study, School of Natural Sciences, Princeton, New Jersey 08540 USA*
[2] *Princeton University Observatory, Princeton, New Jersey 08544 USA*

**Abstract**

The large-scale bulk flow implied by the measurements of distances to brightest cluster members within 15,000 km s$^{-1}$ by Lauer and Postman has the potential to put strong constraints on, or to even rule out, various models of large-scale structure. Using PM simulations 400 $h^{-1}$ Mpc on a side of five cosmological models (Standard CDM, Tilted CDM, HDM, $\Omega = 0.3$ CDM, and PBI), we present detailed Monte-Carlo simulations of the peculiar velocity data, which we use to assess the probability that any given model can explain the data. These models are ruled out at confidence levels ranging from 94% (HDM) to 97% (PBI). However, these results may be affected by the finite size of the simulation volume. If the observed bulk flow remains unchanged when calculated using a refined distance indicator with smaller scatter, all these models will be ruled out at high confidence levels.

## 1 Introduction

As surveys of peculiar velocities of galaxies have become more extensive and probe ever larger scales, it has become increasingly clear that the cosmic velocity field has a very large coherence length [6]. Courteau and collaborators ([4], [7]) presented evidence at this meeting for a bulk flow in an extensive sample reaching to 6000 km s$^{-1}$, while Lauer & Postman ([15]) find a bulk flow at a high statistical significance level in a volume-limited sample of brightest cluster galaxies to 15,000 km s$^{-1}$ from the Local Group. It is the latter observation that we explore in the present paper. Briefly, Lauer & Postman made CCD observations in the $R$ band of 119 Abell ([1]) and ACO ([2]) galaxy clusters, of which 114 have heliocentric redshifts less than 15,000 km s$^{-1}$. Following earlier work of Hoessel ([12]) and Gunn & Oke ([11]), they find a tight quadratic relation between the absolute magnitude of the galaxy within a metric radius of $10h^{-1}$ kpc, and the logarithmic slope of their surface brightness profiles, $\alpha$. Using this as a distance indicator, they measure a bulk flow of the entire sample of $\sim 700$ km s$^{-1}$ as measured in the rest frame of the CMB. This is much larger than would naïvely be expected in any of the currently popular models of structure formation in the universe. However, the observational errors are large, and must be taken into account properly in comparing to models. Feldman ([8]) presents a semi-analytical linear theory treatment of the problem, while here we use Monte-Carlo realizations of the Lauer & Postman data set (hereafter, Warpfire) drawn from $N$-body simulations of various models. We use the latter approach for several reasons: $a$) It allows us to search for subtle systematic effects due to the fact that the distance indicator is calibrated from the same data set from which the bulk flow is measured (cf., the extensive discussion of such effects in [15]); $b$) It is possible that clusters of galaxies exhibit a form of velocity bias by which their velocity field is not well-represented by a volume-weighted sampling of the velocity field as a whole; and $c$) The distribution of distance errors is non-Gaussian, and we are very interested here in the tails of the distribution.

## 2  The $N$-Body Models

The quantity of data that a model for structure formation has to satisfy is impressive. The situation at the moment is unclear, as one group's claim of a measurement which "definitively rules out" a given model is quickly answered by a theorist reviving the model by introducing a new free parameter or physical effect. Thus there currently exist on the market a number of models which are being taken seriously. For the present work, we have examined five of these:

- Standard Cold Dark Matter (CDM), with $\Omega = 1$, $h = 0.5$, and $\sigma_8 = 1.05$;

- Tilted CDM, with $\Omega = 1$, $h = 0.5$, $n = 0.7$, and $\sigma_8 = 0.5$;

- $\Omega = 0.3$-dominated CDM, with $\Omega = 0.3$, $\lambda = 0.7$, $h = 0.65$, and $\sigma_8 = 0.67$;

- Hot Dark Matter (HDM), with $\Omega = 1$, $h = 0.75$, and $\sigma_8 = 1$; and

- Primordial Isocurvature Baryon (PBI), with $\Omega = 0.2$, $x = 0.1$, $m = -0.5$, $h = 0.8$, and $\sigma_8 = 0.9$.

All models (with the exception of HDM) are normalized to the observations of large-scale anisotropy of the COBE satellite ([17]). All models assume Gaussian initial conditions, and all except PBI assume adiabatic fluctuations. Here, $\Omega$ is the cosmological density parameter, $h$ is the Hubble Constant in units of 100 km s$^{-1}$ Mpc$^{-1}$, $\sigma_8$ is the rms matter density fluctuation within $8h^{-1}$ Mpc spheres, $n$ (or $m$ for isocurvature fluctuations) is the spectral index of primordial fluctuations ( $= 1$ if not indicated), $\lambda$ is the contribution of a Cosmological Constant to the space curvature ( $= 0$ if not indicated), and $x$ is the ionization fraction after recombination ( $= 0$ if not indicated). A Particle-Mesh code was used to simulate each model, using $1.56 \times 10^7$ particles within a box of $400 h^{-1}$ Mpc on a side (cf., [18] for details). From each model, the 2000 most massive clusters in the volume are identified as peaks in the density field following [3], and Local Group candidates are identified following [18].

## 3  Monte-Carlo Procedure

We extract Monte-Carlo realizations of the observational sample from a given $N$-body model as follows:

1. We choose an $N$-body point with peculiar velocity smoothed on a $1\,h^{-1}$ Mpc scale of between 520 and 720 km s$^{-1}$, and a local density, smoothed on a $5\,h^{-1}$ Mpc scale, between 0.8 and 2.0, in order to mimic the "Local Group". This particle is taken to be the observer.

2. The "observed" redshift of each cluster $i$ in the box (assuming periodic boundary conditions) is calculated in the "heliocentric" frame:

$$cz_i = H_0 r_i + \hat{\mathbf{r}}_i \cdot (\mathbf{V}_i - \mathbf{V}_0 + \mathbf{V}_B) + error \quad , \tag{1}$$

   where $\mathbf{V}_i$ is the peculiar velocity of the cluster itself, $\hat{\mathbf{r}}_i$ is the unit vector pointing in the direction of the cluster, $\mathbf{V}_0$ is the peculiar velocity of the Local Group candidate, $\mathbf{V}_B$ is a vector of amplitude 300 km s$^{-1}$ and random direction (fixed for a given Local Group), mimicking the correction from the Local Group to heliocentric frame, and $error$ is a Gaussian random variable with standard deviation 185 km s$^{-1}$, which reflects the uncertainty in the measurement of the redshift of a cluster ([15]). The velocity field used is that of the $N$-body simulation, plus a vector (fixed for each Local Group) of random direction whose amplitude is Maxwellian-distributed, with rms value equal to the linear theory expected bulk flow of a box $400\,h^{-1}$ Mpc on a side.

3. The subset of cluster points that have heliocentric redshift less than 15,000 km s$^{-1}$ is noted. A coordinate system is set up in which the Local Group peculiar velocity vector is pointed towards Galactic coordinates (271,29), that is, the direction of the CMB dipole. We reject all clusters that fall closer to the Galactic plane than the limits of the Warpfire sample (defined by drawing limiting lines on the sample as a function of longitude). Galactic extinction causes the cluster sample to be incomplete at higher Galactic latitudes; we model this with the selection function:

$$P(b) = 10^{0.13(1-\csc|b|)} \quad , \tag{2}$$

where the coefficient 0.13 was computed from the Warpfire sample itself using a maximum-likelihood technique ([14] find a much higher coefficient (0.32) for the entire Abell catalog, but extinction effects are not as important for the low-redshift sample we are using here).

4. Of the clusters that remain after this selection procedure, the most massive 114 are chosen as the observational sample. In order to give more weight in the direction of the Shapley Supercluster, where the Warpfire sample shows a dramatic overdensity of clusters, *all* of the 2000 most massive clusters in the box, which fall in the area of sky $300 < l < 360°$, $20 < b < 40°$ are included. In addition, we choose all clusters independent of mass within 7000 km s$^{-1}$, to mimic the fact that the number density of nearby clusters seems to be roughly twice the mean value ([15]). The Local Group redshift of each cluster is noted (Equation 1 without the $\mathbf{V}_B$ term). Each is assigned a logarithmic slope of the surface brightness profile of its brightest cluster member $\alpha$, drawn randomly from the distribution of $\alpha$'s for the Warpfire sample. An absolute magnitude in $R$ within a metric radius of $10h^{-1}$ kpc is calculated using the observed quadratic relation:

$$M_{R,i} = -20.9 - 4.42\alpha_i + 2.76\alpha_i^2 + error \quad , \tag{3}$$

where *error* is a Gaussian-distributed quantity with standard deviation 0.24 magnitudes. The size of the aperture through which the photometry on this brightest cluster member is done depends on the distance assumed for the cluster. Following [15], we tabulate apparent magnitudes assuming the cluster is at its Local Group redshift distance, with an appropriate aperture correction:

$$m_{R,i} = M_{R,i} + 2.5\alpha_i \log_{10}(H_0 c z_{LG,i}/r_i) + 5\log_{10}(r_i/10 \text{ pc}) \quad . \tag{4}$$

At this stage, we have a realization of the Warpfire sample: 114 clusters with positions on the sky, Local Group redshifts, and $\alpha$ and apparent aperture magnitudes for the brightest cluster members.

5. We now follow [15] in order to fit a bulk flow for the data. Our model is that the peculiar velocity field is described by a bulk flow and small-scale velocity dispersion (cf., the discussion in [18]). We place the clusters at the distance indicated by their Local Group redshifts, and fit the data for a quadratic luminosity-$\alpha$ relation as in Equation 3. The residuals from the best-fit relation $\delta M$ correspond to a radial peculiar velocity:

$$V_p = cz_{LG}\left(10^{0.4\delta M/(2-\alpha)} - 1\right) \quad , \tag{5}$$

and a bulk flow is fit to all the data by least squares, where the error for each peculiar velocity is given by $\sigma cz_{LG}/(2-\alpha)$, where $\sigma$ is the rms value of $\delta M$. With the best-fit bulk flow, the distances to each object are updated and apparent and absolute magnitudes recalculated (taking into account the aperture correction), the luminosity-$\alpha$ relation remeasured, and the bulk flow fitted again. This process is repeated until convergence, typically taking four iterations. This bulk flow is that as measured in the Local Group frame; the motion of the Local Group itself (which we assume is known exactly from the measured dipole of the Cosmic Microwave Background) is added to this to give the final dipole in the CMB frame. Unlike [15], we do *not* correct the result for error or geometric bias, as we will compare biased results from the Warpfire sample and from the simulations; this is a valid approach to the extent to which our simulations duplicate the geometry and error distribution of the real data.

6. Steps 1–5 are repeated 500 times for each $N$-body simulation, and statistics are accumulated on the estimated bulk flow in the CMB frame.

## 4 Results, and Comparison with Data

Tod Lauer and Marc Postman have graciously given us a preliminary version of their data set, and we have carried out Step 5 with the real data as well. We find a CMB bulk flow of 773 km s$^{-1}$ towards (355,50) (without error bias or geometric bias correction). This differs slightly from the value reported by [15] because we strictly limit the sample to heliocentric redshift $< 15,000$ km s$^{-1}$ ([15] includes a few clusters at higher redshift associated with the Shapley Supercluster), reducing the sample to 114 clusters. In addition, we assume that the surface brightness profiles are strictly power laws (i.e., $\alpha$ is independent of radius for each cluster), while [15] redoes the aperture photometry from the images themselves on each iteration of their fit. The sky distribution of the Warpfire sample is shown in the upper panel of Figure 1, while one realization of it from the Standard CDM model is shown in the lower panel. The data show much greater clustering than does the realization; it is well-known that CDM systematically underestimates the correlation function of rich clusters [3].

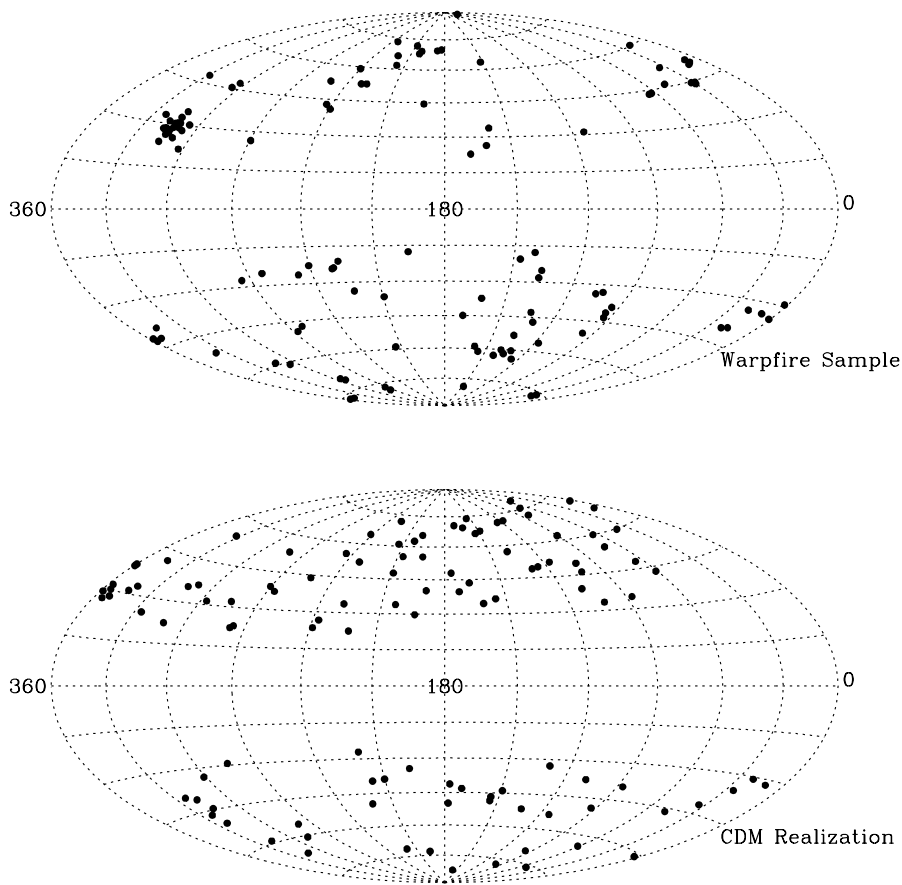

Figure 1: The sky distribution of the Warpfire sample (upper panel) and one realization of the sample drawn from the Standard CDM model (lower panel). Note the far greater clustering apparent in the real data.

The upper panel of Figure 2 shows the results of 500 realizations of the Warpfire bulk flow carried out using the standard CDM simulation. The histogram shows the differential distribution of calculated bulk flows in the CMB frame, while the smooth, monotonically falling curve is the cumulative distribution (on a logarithmic scale!); giving the fraction of realizations which show a bulk flow larger than the value indicated on the ordinate. The vertical dashed line gives the observed bulk flow of 773 km s$^{-1}$. This value is unusual, but not unheard of in a Standard CDM universe; 10.8% of the realizations find a bulk flow larger than this value. Of course, this is not because the bulk flow on these scales is really so large; rather, it is due to the appreciable scatter in the peculiar velocity estimation. Indeed, a scatter plot of the "observed" and true bulk flow within spheres of radius 15,000 km s$^{-1}$ in the standard CDM simulation shows no correlation whatsoever. This becomes clear in Figure 3, which plots the bulk flow velocity estimated from samples of 114 randomly distributed clusters (although still with the selection function of Equation 2 and the Galactic latitude cut), whose peculiar velocities in each component are Gaussian distributed with an rms of 300 km s$^{-1}$, with no bulk flow component. The tail of large derived bulk flows in this case is as extensive as that seen in the case of CDM, meaning that for this data set, an observed bulk flow would have to be appreciably larger than 700 km s$^{-1}$ in order that it be outside the tail due to noise alone. Indeed, one can show in a few steps of algebra that given measured distances for a uniformly distributed full-sky sample of $N$ clusters to radius $R$, with fractional errors in the measured distances $\Delta$, the $1\sigma$ uncertainty in the derived bulk flow is $\sqrt{3/N}\Delta R$. For the present sample ($N = 114, \Delta = 0.16$, and $R = 15,000$ km s$^{-1}$), this gives 390 km s$^{-1}$. An observation of a bulk flow of 773 km s$^{-1}$ is thus significant only at the $2\sigma$ level, by this order-of-magnitude calculation.

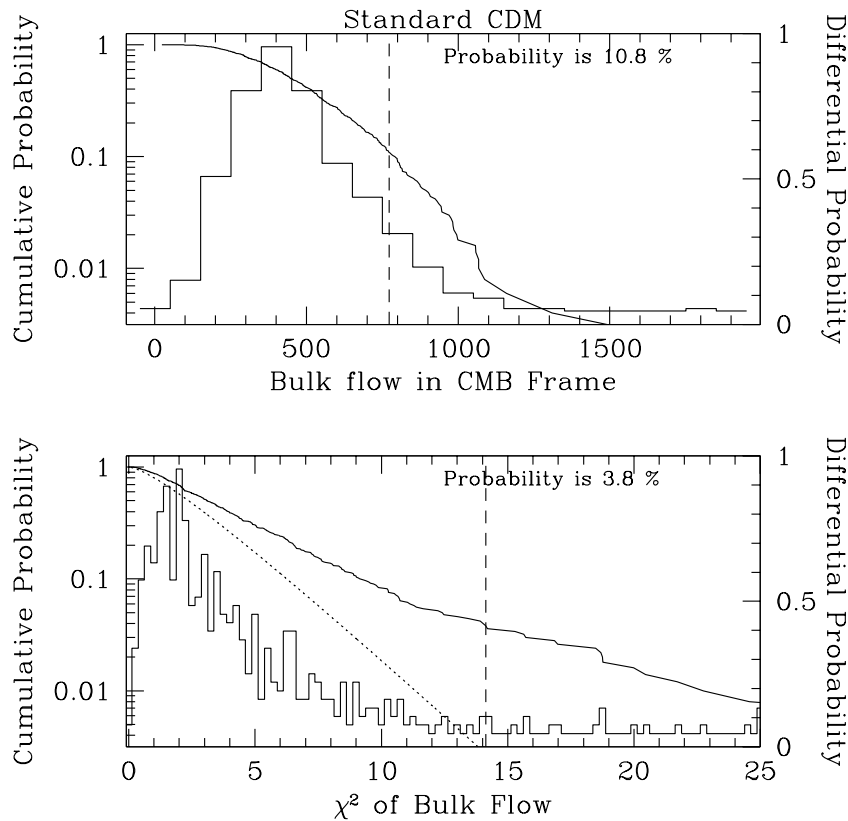

Figure 2: The upper panel gives the differential distribution (histogram) and cumulative distribution (monotonic curve, on a logarithmic scale) of derived bulk flows from realizations of the Warpfire sample from a Standard CDM simulation. The vertical dashed line is the observed bulk flow. The lower panel shows the distributions for the quantity $\chi^2$; the vertical dashed line is the observed value. The dotted line in the expected cumulative distribution of $\chi^2$ with 3 degrees of freedom.

However, the current sample is not isotropic; in particular, because of the excluded region, there is higher sensitivity to detecting the component of the bulk flow towards the Galactic poles than the orthogonal components. This is quantified by the covariance matrix of the bulk flow calculation, whose

inverse is given by:

$$M_{lm} = \sum_{\text{clusters } i} \frac{\hat{r}_{i,l}\hat{r}_{i,m}}{(\Delta r_i)^2} \quad , \tag{6}$$

where the sum is over the clusters in the sample, $\hat{r}_{i,l}$ is the $l$th component of the unit vector towards the $i$th cluster, and $\Delta r_i$ is the error in the measured distance to the galaxy. For the Warpfire sample itself, the eigenvalues of the inverse of $M$ (i.e., the length of the axes of the error ellipsoid) are in ratio $1 : 1.3 : 1.9$, and in fact, the observed bulk flow is directed almost exactly along the *short* axis of the ellipsoid. That is, the statistical significance of the observed bulk flow is much larger than a naïve estimate based on its total amplitude and the isotropic error would indicate. We quantify this by defining a $\chi^2$ statistic of the bulk flow relative to the null hypothesis of no bulk flow at all:

$$\chi^2 = V_l V_m M_{lm} \quad , \tag{7}$$

where $V_l$ is the $l$th component of the derived bulk flow, and Einstein summation is assumed. For the Warpfire sample, $\chi^2 = 14.14$, corresponding roughly to a $3\sigma$ detection given 3 degrees of freedom (the three components of the bulk flow) ([16]). Again, this result differs from that of [15] because the latter authors find a slightly different bulk flow vector, and more importantly, because they derive the covariance matrix by Monte-Carlo methods rather than by Equation 6. For each of the Monte-Carlo simulations, we can calculate the quantity $\chi^2$ for the bulk flow found, using Equation 6 to determine the covariance matrix in each case. The lower panel of Figure 2 shows the results for the Standard CDM simulation. Again, the histogram shows the differential distribution of $\chi^2$, and the monotonic smooth curve shows the cumulative distribution on a logarithmic scale. The dotted curve shows the expected cumulative distribution of $\chi^2$ for three degrees of freedom given the null hypothesis (no bulk flows). The true distribution is more extended, due to the real bulk flows in the models and the non-Gaussian nature of the error distribution (cf., Figure 3; even with no bulk flows, the observed distribution is more extended than the expected $\chi^2$ distribution). The vertical dashed line is drawn at the observed value of $\chi^2 = 14.14$; only 19 realizations out of the 500 (3.8%) show a value of $\chi^2$ larger than that observed.

The $\chi^2$ statistic differs from the amplitude of the bulk flow statistic in the upper panel only to the extent that the error ellipsoid is anisotropic (that is, if it were isotropic, the two statistics would be redundant). Thus it is vitally important that $a$), the errors are modeled correctly in the simulations, and $b$), that the error ellipsoid have the same shape in the simulations as in the real data. To check the first point, we have examined the distributions of the derived scatter of the luminosity-$\alpha$ relation, and found that it indeed has a mean of 0.24 magnitudes (the input value), with an rms of about 0.017. Moreover, the mean rms distance of clusters in the simulations is $8320\,\text{km s}^{-1}$, close to the observed value of $8665\,\text{km s}^{-1}$. As described above, the simulations include the effect of an excluded zone and Galactic extinction, and make an attempt to sample more densely in the direction of the Shapley Supercluster. The distribution of ratios of the largest to smallest axes of the error ellipsoid in the CDM simulations has a mean of 1.85, close to the value of 1.91 seen for the Warpfire sample itself. No correlation is seen between $\chi^2$ and this ratio. Indeed, our statistical results are rather insensitive to exactly how the excluded zones are treated; we experimented with a variety of schemes to match the spatial distribution of the Warpfire sample, and consistently found CDM to be ruled out at the 96% confidence level or better.

[15] correct their calculated bulk flow for error bias (due to the fact that the amplitude of the bulk flow is a positive definite quantity) and for geometric bias due to the fact that the sample is not isotropic. We have corrected neither the observations nor the simulations for these biases, preferring to make direct comparisons of uncorrected quantities. This is a valid procedure to the extent that the biases (which depend on the distribution of clusters and the scatter in the distance indicator relation) are the same for both models and data. In any case, the error bias is *not* subtracted by [15] for the $\chi^2$ statistic, and the geometric biases are small. Moreover, calculations of the geometric biases from the simulations are in good agreement with those listed in [15].

Comparison of Figures 2 and 3 indicates that any model which exhibits intrinsic bulk flows on the scales probed by the Warpfire sample which are small relative to the noise will be ruled out at a similar confidence level by the Warpfire observation. Our results for all models are shown in Table 1, listing the fraction of realizations in each model which gave bulk flows and $\chi^2$ values larger than those found in the real data. The second column of the table lists the actual bulk flow within the spheres (i.e., without any errors applied) found by fitting to the radial components, and averaged over realizations; as we discuss in the following section, this gives different results from summing up all three components of the peculiar velocity in the presence of flows on intermediate scales. Next follows the mean bulk flow over realizations, now with the errors applied. The fraction of realizations giving bulk flows and $\chi^2$ values larger than those observed follow in the next two columns. All five models are ruled out at better than the 94% confidence level by the $\chi^2$ statistic, with HDM faring the best, and PBI faring the worst (97%). These results are disappointing, in that we are in the all-too-familiar situation of models being "ruled out" at only the $\sim 2\sigma$ level, which is not strong enough to be definitive. However, note that all models fare roughly equally as well; the differences in true bulk flows between models is lost in the noise.

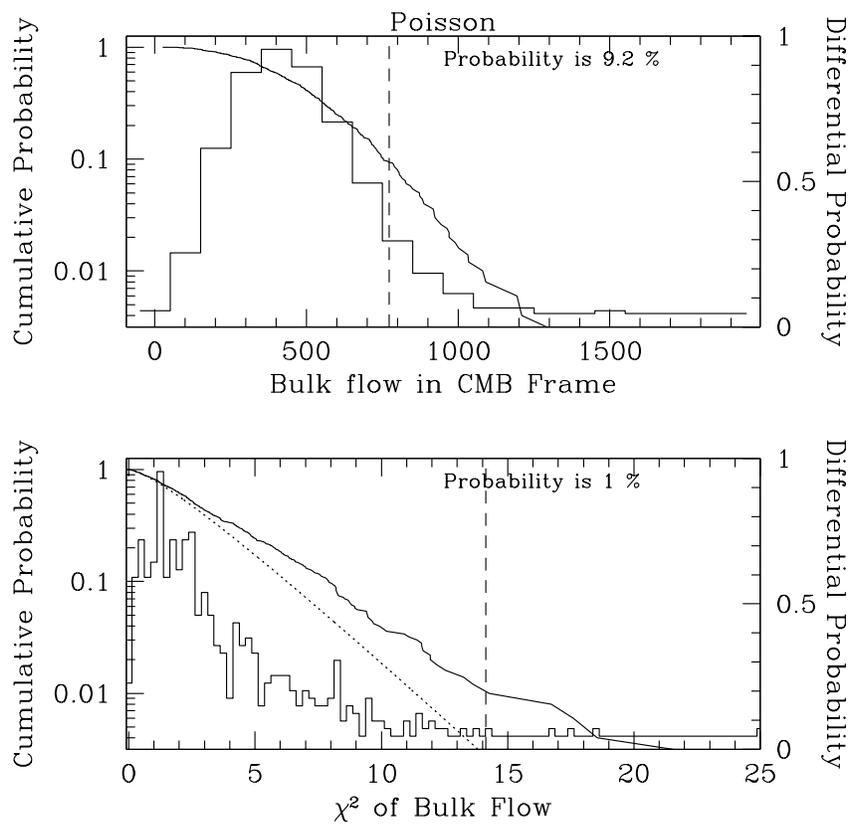

Figure 3: As in Figure 2, for a Poisson distribution.

Finally, we did two further experiments: cutting the observational sample at $|b| = 30°$, to minimize the effects of Galactic extinction (leaving 86 clusters), and cutting at $cz = 10{,}000$ km s$^{-1}$, to increase the signal-to-noise ratio of the bulk flow (leaving 43 clusters). In the former case, $\chi^2$ drops to 7.95, and 12.6% of the Standard CDM simulations can match this value. In the latter case, $\chi^2$ stays high at 13.81, but the much smaller number of clusters increases the noise of the estimate, and 5.2% of the realizations are able to match the observed value.

Table 1

| Model | True Bulk Flow km s$^{-1}$ | Measured Bulk Flow km s$^{-1}$ | % exceeding Bulk Flow | % exceeding $\chi^2$ |
|---|---|---|---|---|
| Standard CDM | $132 \pm 56$ | $484 \pm 222$ | 10.8% | 3.8% |
| Tilted CDM | $166 \pm 64$ | $493 \pm 216$ | 10.6% | 4.2% |
| $\Omega = 0.3$ CDM | $176 \pm 79$ | $490 \pm 240$ | 13.4% | 4.6% |
| HDM | $139 \pm 56$ | $502 \pm 231$ | 12.4% | 5.8% |
| PBI | $123 \pm 59$ | $422 \pm 203$ | 5.6% | 2.6% |
| Poisson | — | $471 \pm 217$ | 9.2% | 1.0% |
| Standard CDM[a] | $126 \pm 54$ | $563 \pm 253$ | 30.2% | 12.6% |
| Standard CDM[b] | $189 \pm 80$ | $522 \pm 228$ | 15.2% | 5.2% |

[a] Cut at $|b| = 30°$.
[b] Cut at $cz = 10{,}000$ km s$^{-1}$.

## 5  Discussion

The velocity field model which we have assumed in fitting bulk flows to the data and the simulations consists of a constant large-scale flow plus small-scale random velocities which give rise to the scatter in the luminosity-$\alpha$ diagram. In fact, the velocity field in general has components on all scales, which means that the bulk flow derived from a data set may not be equal to our intuitive notion of what it should be. Imagine an observer in a homogeneous universe, and add a single cluster (not at the origin), which gravitationally attracts the particles around it. The mean bulk flow of a sphere centered on the observer, which contains the cluster, is non-zero (and indeed is proportional to the distance between the observer and the cluster), despite the fact that there is no source of gravity from outside the sphere ([13]). Moreover, one can show that in this case, the bulk flow found by fitting our model to the radial components of the peculiar velocities is quite different from that calculated by summing up the peculiar velocity vectors (if we could somehow observe the transverse components). One could get around the latter problem if one had *a priori* knowledge of the velocity correlations on different scales (in the sense of [10], for example), and included this in the least-squares solution for the bulk flow using radial peculiar velocities alone. For the Warpfire sample, we have seen that random errors due to the finite scatter of the distance indicator relation dominate completely, and the refinements of including a model for the velocity correlations are unlikely to make a difference. This however is an avenue we will pursue with peculiar velocity data sets at lower redshifts, with higher signal-to-noise ratio per point.

There are a number of ways in which the analysis discussed in this paper might be improved. Perhaps most obvious is the size of our simulation volume; a $400h^{-1}$ Mpc cube contains only 4.5 independent spheres of radius 15,000 km s$^{-1}$, a rather small number. We have argued that because the expected bulk flow in any of the models is small, the measured bulk flow is much more a function of the errors than of the velocity field itself. Moreover, each realization of the data set contains a different realization of the power on scales larger than the box size; that is, we have accounted for power on scales larger than the box size. However, we can see the largest waves directly in the (biased) cluster distribution, especially in the Tilted CDM and $\Omega = 0.3$ CDM models, which is having a noticeable effect on our statistics. We are currently doing a series of PM simulations within cubes $800h^{-1}$ Mpc on a side to rectify this; preliminary results are consistent with those quoted in this paper, and details will be reported in due course.

In [18], we made comparisons of the bulk flow found in various peculiar velocity data samples with the small-scale velocity dispersion; the ratio of the two, the Cosmic Mach Number, gives a measure of the relative amounts of large-scale and small-scale power. The Warpfire sample does not allow a straightforward measure of the small-scale velocity dispersion, as the distance indicator relation is calibrated directly from the data set itself; we have no direct way of distinguishing intrinsic scatter in the relation from the effects of peculiar velocities. However, the errors in peculiar velocities due to the former scale proportional to the distance of each cluster, while the latter are nominally independent of distance. [15] used this fact to estimate that the small-scale velocity dispersion was less than 250 km s$^{-1}$ (1-d). It would be instructive to make a direct comparison of this statistic between the data and the models.

Given the fact that the Warpfire data set appears to rule out so many models "in one fell swoop" (albeit at the $\sim 2\sigma$ level), one might be tempted to explain their bulk flow away to some subtle systematic effect. It should be pointed out, however, that the Warpfire sample is selected using well-defined criteria, covers the entire sky, has uniform quality data and is self-calibrating. No other peculiar velocity survey can boast all of these qualities. Moreover, Lauer & Postman are extending the sample in two ways. First, they have discovered that velocity dispersion is a second parameter in the luminosity-$\alpha$ relationship, and that including velocity dispersion decreases the scatter of the distance indicator relation from 0.24 to $\sim 0.18$ mag. We have carried out Monte-Carlo simulations of the Warpfire sample using this smaller scatter; if their bulk flow result holds up, all of the models discussed here will be ruled out at better than the 1% confidence level. This is not surprising; it is the errors which drive the tail of the distributions seen in Figure 2, and the realizations with the largest "measured" bulk flows do not have unusually large "true" bulk flows. Decreasing the distance indicator errors will greatly decrease the extent of the tail. Second, they have started a program to obtain distances to the brightest cluster members of a complete sample of Abell clusters to 24,000 km s$^{-1}$, greatly increasing the effective volume of the survey. We eagerly look forward to the results of these observations.

There now exist data on a very large range of scales which put important constraints on the cosmic velocity fields. On the largest scales, the COBE measurements of anisotropy, if interpreted as due to the Sachs-Wolfe effect, constrain peculiar velocities on scales above $\sim 600\,h^{-1}$ Mpc. The Warpfire sample probes scales of $\sim 100\,h^{-1}$ Mpc, while the bulk flow reported at this meeting by [4] is sensitive to power on scales roughly half of this. Various data sets have measured bulk flows on scales of 10–20 $h^{-1}$ Mpc (see [18] for a discussion), while [5] and more recently [9] have accurately measured the velocity dispersion at $1\,h^{-1}$ Mpc. We have seen in the current work that all of the currently popular models which we have tested are ruled out at roughly the $2\sigma$ level. We are working now to find a phenomenological model that will fit all available data.

**Acknowledgements.** We are grateful to Tod Lauer and Marc Postman who have given us the Warpfire data in advance of publication, and who have cheerfully responded to our many questions.

MAS is supported at the IAS under NSF grant PHY92-45317, and grants from the W.M. Keck Foundation and the Ambrose Monell Foundation. RYC and JPO are supported by NASA grants NAGW-765 and NAGW-2448, and NSF grants AST90-06958 and AST91-08103.